# A Mechanism-Guided Inverse Engineering Framework to Unlock Design Principles of H-Bonded Organic Frameworks for Gas Separation

Yong Qiu[1], Lei Wang[1], Letian Chen[2], Yun Tian[1*], Zhen Zhou[1,2*], Jianzhong Wu[3]

[1]Interdisciplinary Research Center for Sustainable Energy Science and Engineering (IRC4SE[2]), School of Chemical Engineering, Zhengzhou University, Zhengzhou 450001, Henan, China
[2]School of Materials Science and Engineering, Institute of New Energy Material Chemistry, Renewable Energy Conversion and Storage Center (ReCast), Key Laboratory of Advanced Energy Materials Chemistry (Ministry of Education), Nankai University, Tianjin 300350, China
[3]Department of Chemical and Environmental Engineering, University of California, Riverside, CA, 92521, USA
* Corresponding author. E-mail: ytian009@zzu.edu.cn; zhouzhen@nankai.edu.cn

## Abstract

The diverse combinations of novel building blocks offer a vast design space for hydrogen-boned frameworks (HOFs), rendering it a great promise for gas separation and purification. However, the underlying separation mechanism facilitated by their unique hydrogen-bond networks has not yet been fully understood. In this work, a comprehensive understanding of the separation mechanisms was achieved through an iterative data-driven inverse engineering approach established upon a hypothetical HOF database possessing nearly 110,000 structures created by a material genomics method. Leveraging a simple yet universal feature extracted from hydrogen bonding information with unambiguous physical meanings, the entire design space was exploited to rapidly identify the optimization route towards novel HOF structures with superior Xe/Kr separation performance (selectivity $>10^3$). This work not only provides the first large-scale HOF database, but also demonstrates the enhanced machine learning interpretability of our model-driven iterative inverse design framework, offering new insights into the rational design of nanoporous materials for gas separation.

## Introduction

The efficient adsorption and separation of xenon (Xe) and krypton (Kr) hold substantial industrial significance due to the specific applications in extensive applications in nuclear industry, space exploration, commercial lighting and medical devices[1-4]. However, due to the nearly identical kinetic diameters (Kr = 3.66 Å; Xe = 4.05 Å), ultra-low concentrations (Kr = 1.14 ppmv; Xe = 0.09 ppmv), and extremely low polarizabilities, the major challenges associated with the separation of Xe and Kr remain significant[5]. Traditionally, cryogenic distillation is the most mature technology to separate pure Xe and Kr (20/80 (v/v)) from air, but these processes are energy- and capital-intensive[6]. Metal-organic frameworks (MOFs) and covalent organic frameworks (COFs) have been the most developed porous materials for Xe/Kr separation, yet their commercialization is significantly hindered by issues related to material stability, high costs, and environmental pollution. Therefore, exploring stable, environmentally friendly porous adsorbents with controllable structures to balance capacity and selectivity remains both appealing and challenging[7].

Compared with other classes of nanoporous materials dominated by relatively stronger intra-molecular interactions like MOFs and COFs, Hydrogen-bonded organic frameworks (HOFs) possess unique advantages, such as milder synthesis conditions, facile solution processability and recyclability, which are usually attributed to their weaker inter-molecular interactions and higher reversibility of hydrogen bonding[5]. In addition, previous studies have demonstrated that HOFs exhibit significant potential for Xe/Kr separation, with excellent thermal and chemical stability, and can be efficiently recovered and reused through simple recrystallization, thereby greatly reducing the operational cost of the adsorbents[1, 7].

To date the total number of experimentally reported HOF structures is still limited to $10^2$, primarily owing to the difficulty in materials synthesis[8, 9]. Although the separation of Xe/Kr in MOFs[10, 11] was attributed to modified polarizability of guest molecules induced by designed functional groups, the separation mechanism for Xe/Kr in HOF materials remains unknown, which severely hindered the rational design of high-performance HOF materials. With large unexplored design space, the recent application of modular design in HOF synthesis[12, 13], and the emergence of automated laboratories[14-16], in-silico design of HOFs is highly attractive as a viable alternative to drive the fundamental understanding of the separation mechanism and accelerate the discovery of targeted HOFs.

As a well-established high-throughput computational method to predict gas adsorption and separation, Classical Density Functional Theory (CDFT) has demonstrated its outstanding efficiency and accuracy under various scenarios[17-19]. Integration of machine learning techniques with CDFT has further improved the computational efficiency and revolutionized the design strategy in recent years[20, 21]. In our previous work, forward design strategy was fully addressed to identify high-performance gas separation materials from an existing COF database[22]. Given the vast chemical space with exponentially increasing building units, it is highly demanding to incorporate inverse design approaches to uncover the optimization route for novel HOF structures[23-25]. Regrettably, the applicability of current inverse design methods is usually limited to their initial dataset or their analogues, severely hinders the iteration process of new structures[26-28].

In this work, we developed a hypothetical framework to construct a large-scale HOF database with a genomics-based method as shown in Figure 1, enabling the rapid generation of HOF structures and high-throughput predictions of gas separation performance of Xe/Kr by CDFT. Through incorporation of generic algorithm with active learning, for the first time, we established an iterative inverse design framework to efficiently optimize the structure of novel HOF materials to identify the target ones. As an illustration, we successfully designed a novel HOF structure with Xe/Kr selectivity over $10^3$, exceeding the best existing adsorbent for Xe/Kr separation[1, 29, 30]. Furthermore, after comprehensive analysis with our framework, it is conclusively demonstrated that the Xe/Kr separation mechanism in HOFs is not primarily governed by polarizability differences. Instead, the separation process is driven by a synergistic dual mechanism involving pore sieving effect and the variable hydrogen bonding intensity. More excitingly, the fundamental mechanisms driving Xe/Kr separation was utilized to establish a novel descriptor with enhanced interpretability and universality, which was constructed from both the hydrogen-bond networks and key structural features of the materials. With the iterative inverse engineering framework, we expect to provide a new paradigm for developing nanoporous materials with materials genomics strategies, offering guiding principles for the experimental design of HOFs.

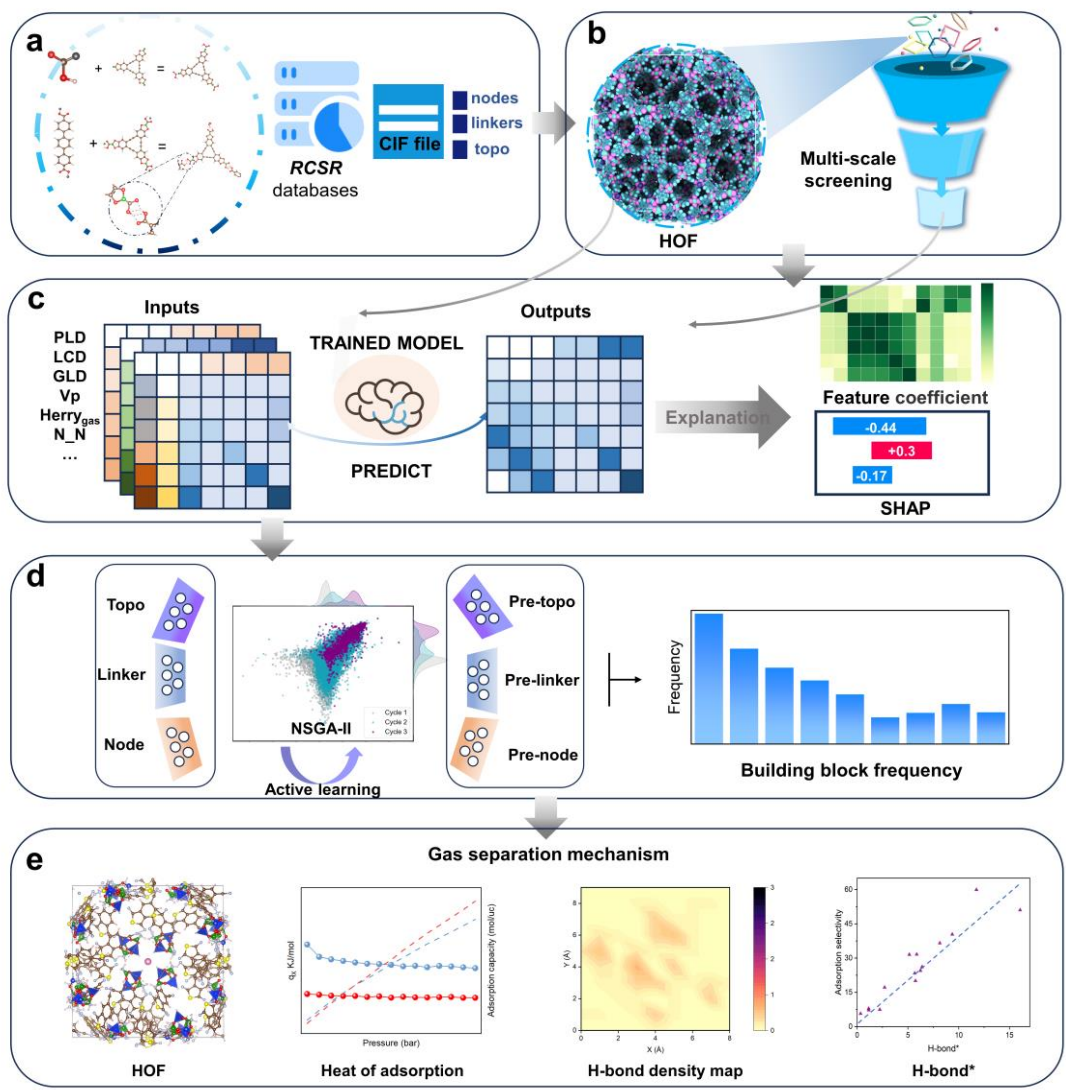

**Figure 1.** Workflow of mechanism exploration based on a self-built database and the iterative inverse engineering framework. a. In-silico construction of a HOF database. b. Feature extraction and high-throughput prediction of separation performance using CDFT. c. Machine learning model training and interpretability analysis. d. Active learning promoted iterative inverse design of HOFs. e. Mechanism revelation of Xe/Kr separation in HOFs.

## Results

### Materials genomics-based algorithms for HOF construction

The primary distinction between HOF, COF, and MOF materials lies in the differences of the bonding interactions between the building blocks (BBs)[31, 32]. Specifically, when the building units possess functional groups capable of acting as hydrogen-bond donors or acceptors, thereby facilitating the formation of stable hydrogen-bond networks, it becomes theoretically possible to construct novel HOF structures based on existing topological frameworks. Taking the building units and topological structures of COF materials collected in the Reticular Chemistry Structure Resource (RCSR)[33] database as a template, we propose a methodology to construct BBs for hypothetical HOFs by functionalizing the building units (as shown in Figure 1a). Given the functional groups such as -OH, -COOH, -$NH_2$, and -F serving as hydrogen-bond donors or acceptors to facilitate the formation of diverse hydrogen-bond networks, modified BBs were produced to constitute novel HOF structures, which finally realized the transformation from COF to HOF as inspired by experimental evidence. Through the incorporation of these functional groups, the BBs were not only saturated but also capable to form hydrogen-bond networks, thus satisfying the preliminary criteria for the design and synthesis of HOFs. The names of these new BBs are derived by adding the functional group name to the original BB name. For example, an original BB named 'C6' would be modified by adding a COOH functional group, resulting in 'C6_COOH'.

The detailed process for constructing HOFs is as follows: First, a gene library for HOFs is established, where each HOF name includes topology + node + edge ligand. After assessing the degree of compatibility, BBs and topological units are assembled using pormake[34] to construct complete new HOFs. Furthermore, through a parallel design approach, we significantly accelerated the materials design process and established the first theoretical HOF database, comprising nearly 110,000 distinct HOF structures. Moreover, this approach offers sufficient versatility and flexibility for users to perform targeted materials design, enabling facile modification of functional groups and integration of diverse BBs to construct materials tailored to specific applications.

**Theoretical HOF materials database**

Under typical industrial separation conditions (296 K, 1 bar, 20/80 molar ratio for Xe and Kr)[1, 35], we performed high-throughput CDFT calculations for Xe/Kr gas separation on the constructed HOF database after excluding those with inaccessible pore geometries for gas molecules, where detailed comparison and validation with available experimental data are shown in Figure S1. Key structural characteristics including Largest included sphere along free sphere path diameter (PLD), Largest free sphere diameter (LCD), pore volume, framework density, pore characteristics, specific surface area, etc., and chemical features including element ratios, unsaturation, electronegativity ratio, etc., were calculated for subsequent machine learning implementations.

The selectivity values in our database range from 0 to $10^4$, reflecting the diversity of our comprehensive dataset, with the top-performing materials significantly surpassing the separation capabilities of existing experimental HOF materials. Detailed data distributions of the entire database can be found in Figures S2 to S11. For analytical clarity, we categorized the HOFs in the database into three groups based on their Xe/Kr adsorption selectivity: Class 1 (top 10%), Class 2 (top 10%-30%), and Class 3 (the remaining materials). As illustrated in Figure 2, an apparent correlation can be drawn between the Xe/Kr separation performance and the physical structure of these materials. Materials exhibiting superior separation performance typically feature relatively smaller pore sizes with PLD around 15 Å (Figure 2a), higher framework densities (Figure 2b), and smaller void fractions (Figure 2c), which can be attributed to the inert nature of Xe and Kr and the dominating size-exclusion effects. In terms of these geometric features, the void fraction plays a vital role in measuring the upper limit of capacity for gas molecule accommodation, and with well-defined pore size and density of frameworks, the optimal HOF is capable to effectively differentiate guest molecules through non-uniform intermolecular interactions with its hydrogen-bond networks, leading to enhanced separation efficiency. Meanwhile, Henry's coefficient, which accounts for the thermodynamic interactions between the guest and host molecules, also determines the relative adsorption strength and separation performance of porous materials. As shown in Figure 2d, Henry's coefficients for Xe/Kr in HOFs exhibit a consistent upward trend, regardless of whether they are evaluated for ideal adsorption selectivity (Figure 2d) or selectivity under practical operating conditions (Figure 2e), indicating that separation performance might be primarily driven by the difference in interacting affinity with frameworks. However, depending solely on any single physiochemical character is inadequate to effectively distinguish the performance of HOFs.

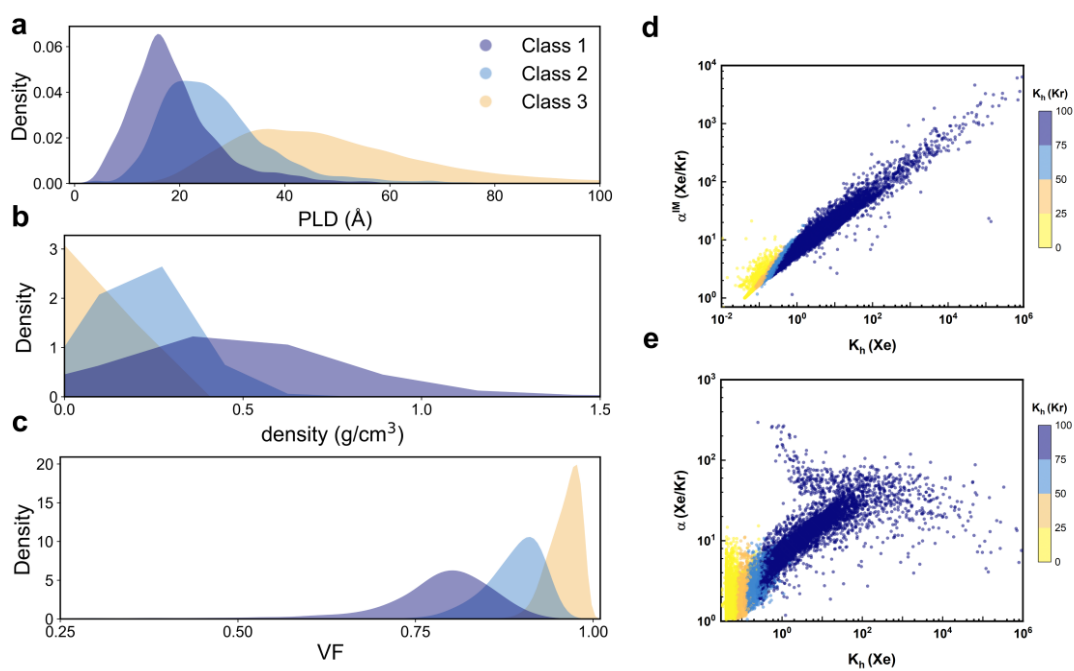

**Figure 2.** Relationship between HOF framework characteristics and Xe/Kr adsorption selectivity. Density plots of the framework's physical structure and adsorption selectivity: PLD (a), density of frameworks (b), and void fraction (c), as well as the relationship between Henry's constant and ideal adsorption selectivity (d) and practical adsorption selectivity (e). In panels (d) and (e), the color map represents the variation in the Henry's constant for Kr, with darker colors indicating higher Henry's coefficients.

While experimental results show the influence of measurable physicochemical properties on performance, high-throughput calculations and machine learning models enable rapid prediction of materials performance and offer deeper insights into the underlying mechanisms. In this regard, we conducted feature selection using the Pearson correlation coefficient (Figure 3a) integrated with unsupervised learning methods, including t-SNE and PCA (Figures S12-S14). Our analysis revealed that the physical structural features predominantly influence the target property prediction, while the chemical characteristics exhibit a relatively weaker correlation with the separation performance. Owing to the inert nature of noble gases, Kr and Xe molecules are less prone to chemically react with the framework and are predominantly adsorbed through physical interactions. To more specifically quantify the relationship between features and performances, we applied a variety of machine learning models (listed in Table S1) to predict the actual Xe/Kr adsorption selectivity of the HOFs (Figures S15-S16). Through hyperparameter optimization for each model, with an 80:20 training-to-test set ratio and 5-fold cross-validation, we identified the model with the best performance. Our results show that tree-based models, such as Random Forest (RF) and Gradient Boosting Regression (GBR), outperformed those linear regression algorithms including Least Absolute Shrinkage and Selection Operator (LASSO), Linear Support Vector Machine (Linear SVM), and Partial Least Squares (PLS) regarding data regression. Neural-network-based models like Artificial Neural Network (ANN) and K-Nearest Neighbors (KNN) were prone to overfitting, indicating excessive model complexity. Among all of them, the GBR model delivered the best performance, achieving an $R^2$ of 0.905 on the test set as illustrated in Figure 3b. In spite of the minor divergency in regions of high selectivity due to sparse data distribution in these areas, the machine learning model generally provides satisfactory predictions for separation performance based on physical properties. To further investigate the structure-property relationships of the materials, we conducted SHapley Additive exPlanations (SHAP) analysis on the trained GBR model to assess the significance of each feature, as depicted in Figure 3c and 3d. The top nine most important features, all of which are physical properties, are listed, based on both the average SHAP values and individual feature contributions, which further substantiate our earlier feature selection analysis, confirming that physical properties are most critical in determining separation performance, with framework density ranking among the top three most significant features. However, our analysis of individual feature weights (Figure 3d) reveals the coexistence of both positive and negative correlations among features across different frameworks. This observation indicates that these descriptors cannot be used in isolation to reliably assess materials performance, highlighting the need to uncover the more complex underlying mechanisms that govern the inter-correlation of these individual features.

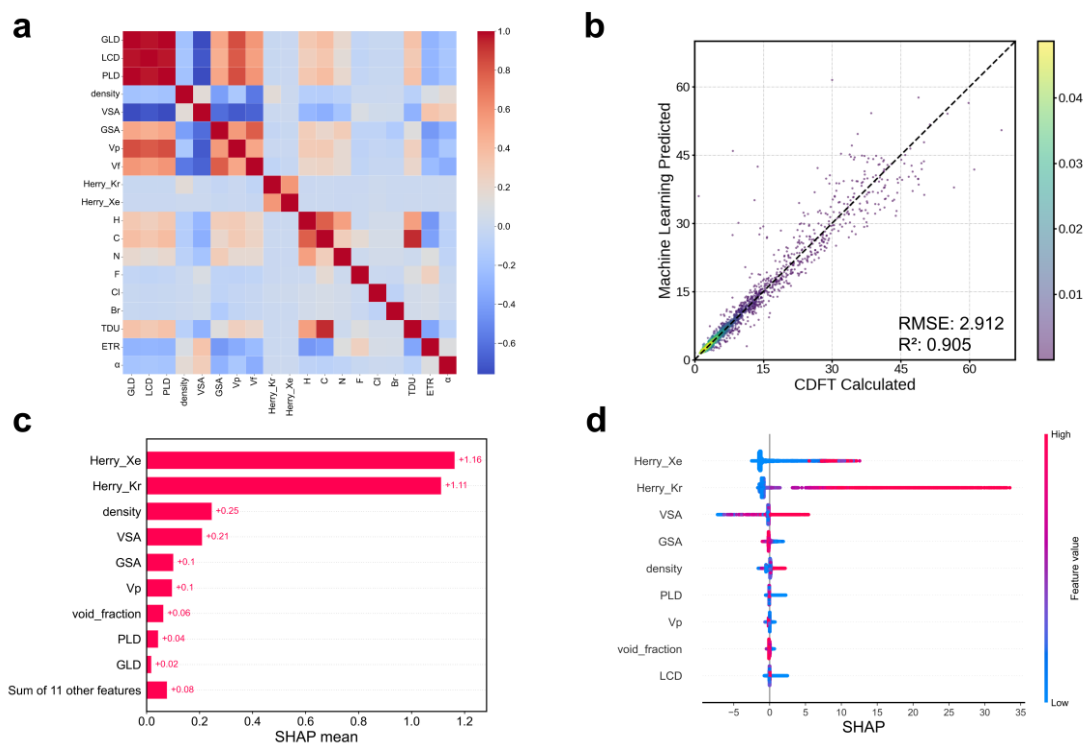

**Figure 3.** Correlation of different features, importance analysis, and prediction performance. a. Pearson correlation coefficients, with color mapping representing both the direction (positive/negative) and magnitude of correlations (The top 10 are physical features, while the others are chemical features. The full names are provided in Table S1.); b. GBR machine learning model performance on the test set; c. Average SHAP values, indicating the mean feature importance across all samples; d. SHAP beeswarm plot visualizing the distribution of SHAP values, with higher values represented by redder colors.

## Performance-driven inverse design of HOFs

Inverse design represents a paradigm shift in materials discovery, wherein the process initiates from the desired materials performance and systematically traces back to the requisite structural properties. This approach facilitates more efficient and targeted materials development. However, materials properties are typically governed by multiple performance factors rather than a single determinant. An exclusive focus on a singular objective during the design process may lead to impractical outcomes, such as the materials with extremely high selectivity while very low adsorption capacity. To address this issue, we employed the NSGA[36] (Non-dominated Sorting Genetic Algorithm) for multi-objective optimization, incorporating both the adsorption capacity and selectivity for the target gas species of Xe. The simultaneous achievement of high adsorption density and high selectivity is set as a key indicator of superior materials performance, demonstrating both efficient molecular uptake capacity and precise target-specific recognition capabilities. Given that inverse design is intrinsically associated with the BBs and topology of the material, we incorporate these critical factors as high-dimensional input features. Leveraging the pre-trained MOF-NET[34] model, we predict Xe/Kr adsorption selectivity and Xe adsorption capacity of HOFs, enabling a more targeted and efficient design process. The model employs a resampling strategy to address data imbalance by selectively augmenting underrepresented regions within the dataset, thereby enhancing the overall balance and representativeness of the database. As shown in Figure S17, the R² values for the test set consistently approximate 0.99, demonstrating significantly improved predictive accuracy compared with models trained exclusively on physical and chemical features as seen in Figure S15. The enhanced performance can be exclusively attributed to the model's implicit incorporation of critical structural factors, particularly the coordination environment of diverse BBs, which encompasses the synergistic effects of hydrogen-bond networks and their surrounding chemical environments.

Compared with Deng and Sarkisov's multi-objective optimization model[37], our approach represent a substantial advancement through integration of an active learning process into the NSGA-II framework, thereby enhancing the model's adaptive capabilities and optimization efficiency. By selecting the top 10% materials predicted with the most promising performance in each generation for further materials design and high-throughput computation, we continuously assess the model's accuracy and facilitate iterative updates, which are subsequently applied to the next cycle of structure predictions, thereby creating a self-improving optimization loop. This enhanced sampling methodology effectively mitigates decision-making errors arising from materials performance deviations from the pre-trained database, while simultaneously accelerating the discovery and optimization of next-generation materials. Additionally, NSGA-II optimizes the selection process by incorporating crowding distance and elitism, which enhances multi-objective optimization, maintains population diversity, and reduces premature convergence. In this work, as a proof of concept, we implemented three genetic cycles, with each cycle generating approximately 100,000 potential structures (note that duplicate HOFs are systematically eliminated in subsequent generations to maintain structural diversity). As shown in Figure 4a, the materials generation and performance prediction for the three cycles are depicted. Through comprehensive analysis of the performance density distribution across successive generations, it is evident that the inverse design process effectively directs the transition from performance-oriented to structure-based materials design. In the newly generated third-generation materials, the majority exhibit superior Xe gas density and adsorption selectivity. Notably, when compared with the randomly generated ~$10^5$ structures, the proportion of materials with excellent adsorption performance increased substantially. These results underscore the effectiveness of our inverse design approach in generating a higher proportion of candidates with optimized materials performance. Building upon these findings, we conducted a detailed frequency analysis of the BBs present in the third-generation materials to identify key structural motifs contributing to these performance enhancements. Figure 4b and 4c present the top-ranked node and edge building blocks, respectively. It reveals that -F, -O, and -N functional groups, prone to form strong hydrogen bonding, dominate the BBs composition, which is crucial for facilitating efficient Xe/Kr gas separation. Furthermore, we observed a significantly higher frequency of the C36_COOH building block compared with other node types, which can be attributed to C36_COOH's unique structural characteristics in matching more edge building blocks, consequently promoting the formation of additional hydrogen bonds with increased framework density, as illustrated in Figure S16. To quantitatively evaluate this structural distinction, we conducted a comparative analysis of hydrogen-bond networks across nearly 50 structure variants, in which only the node type was displaced (replacing typical high-frequency-node C36_COOH with low-frequency node

C36_NH3). As demonstrated in Figure S18, the HOF structures with C36_COOH node consistently exhibit more hydrogen bonds in virtually all cases, which can be attributed to the presence of additional '-O' groups in C36_COOH, which serve as effective hydrogen-bond acceptors and intensify the hydrogen-bond networks with edge components, thereby enhancing adsorption and separation capabilities.

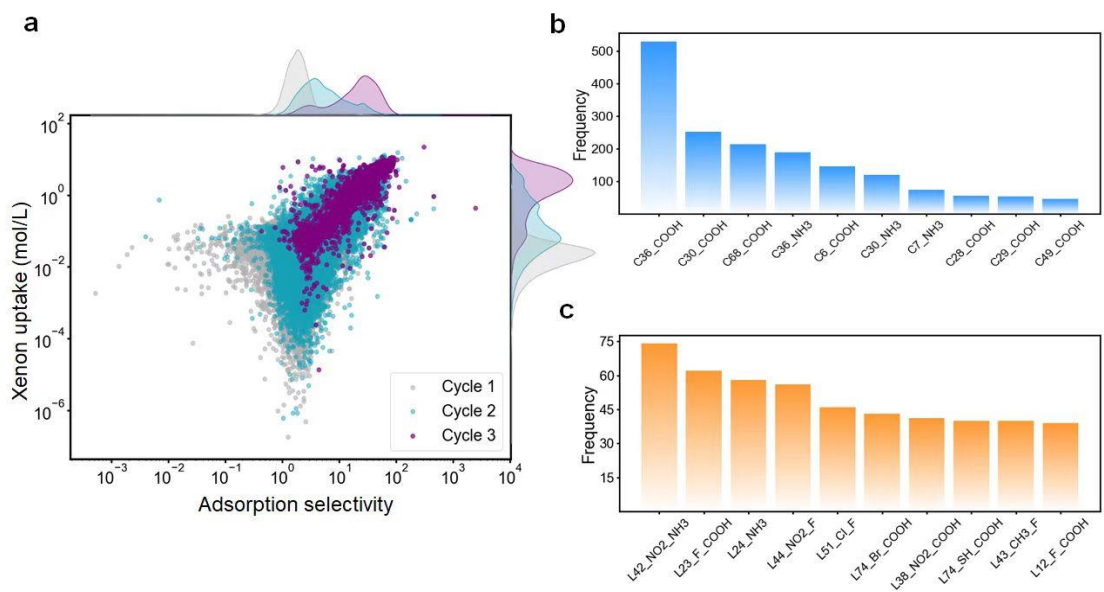

**Figure 4.** Performance analysis of inverse-designed materials. a. Inverse design performance over three cycles, with shading from light to dark indicating an increase in iteration cycles. Higher values correspond to better performance. The materials density analysis for the two performances metrics is shown at the top and right sides. Frequency analysis of the top-ranked node (b) and edge (c) building blocks for materials in cycle 3.

## The separation mechanism of Xe/Kr in HOFs

To gain deeper insights into the adsorption mechanisms of guest molecules within the framework, we take 'sxt+C6_NH3+L39_CN_F' as a representative study (Figure S19). As shown in Figure 5a and 5b, we first determined the adsorption sites of Xe/Kr gas molecules in the HOF based on the adsorption density map (The XY direction is shown in Figure S20a). The adsorption sites for Xe and Kr are essentially the same, while the adsorption intensity of Xe is stronger than that of Kr, as evidenced by the more concentrated adsorption pattern. As shown in the adsorption density map, the central cage-like region serves as an important adsorption site within the framework. In addition, heat of adsorption exhibited a consistent decreasing trend with increasing loading over a wide pressure range shown in Figure 5e, reflecting that the dominating molecular interactions transit from guest-host to guest-guest, and larger initial adsorption heat for Xe indicates stronger intermolecular interaction between Xe and the framework.

To understand the adsorption separation mechanism facilitated by the unique hydrogen-bond networks, we also investigated the adsorption sites within corresponding COF structures with the same topology for direct comparison. It was observed that the corresponding COF framework adsorbs guest molecules preferentially at the edge sites (Figure S20a), instead of the central cage-like region as promoted by the hydrogen-bond networks in HOFs (Figure S20b). From the adsorption energy analysis (Figure S21) and the differential charge distribution (Figure S22), the adsorption strength for Xe at this site is higher than that for Kr and the electron transfer between Xe/Kr and the surrounding BBs is nearly zero, indicating negligible polarizability difference, consistent with non-polar nature of inertia gases. Rather than the polarizable effect[38], the primary factor driving Xe/Kr separation is probably the difference in the intermolecular interactions between the hydrogen-bond acceptors and gas molecules, which is closely related to the intensity of hydrogen-bond networks within the system. Therefore, we statistically analyzed the distribution of hydrogen bonding within the framework, as shown in Figure 5c. It was found that this distribution aligns well with the adsorbed density distribution of guest molecules, suggesting that the arrangement of hydrogen-bond networks reflects the relative strength of guest molecule adsorption within the framework. In contrast, there is no such apparent correlation in corresponding COF structures (Figure 5d), which demonstrates the fundamental difference in the adsorption separation mechanisms between these two types of framework materials.

Is there a universal correlation between the hydrogen-bond networks and the gas separation performance of HOF materials? To address this question, we randomly selected approximately 500 HOF structures and examined the direct relationship between the number of hydrogen bonds and adsorption selectivity. Through Lasso regression, it was found that the number of hydrogen bonds consistently ranked among the top four features with the highest weight (Figure S23). However, the number of hydrogen bonds itself does not exhibit strong correlation with adsorption selectivity as seen in Figure S24. Taking the top-ranking structural features into consideration, density of the framework was further identified to possess stronger correlation with adsorption selectivity as shown in Figure S25. As well as the void fraction and the simplest yet straightforward feature, it does not involve any complex calculations like Henry's constants. Through integration of key structural factors with the information of hydrogen-bond networks, we thus propose a novel descriptor, H-bond*, to represent the effective hydrogen bonding density, which effectively captures the linear correlation with adsorption selectivity without any input from molecular simulation data as illustrated in Figure 5f. It is worth noting that the slight deviation from linearity is probably due to the varied interaction strength between guest molecules and hydrogen-bond acceptors, which is acceptable as a universal screening tool for rapid performance assessment that bypasses cumbersome molecular simulations. Additionally, the enhanced linear correlation was also confirmed with symbolic regression using SISSO as seen in Figure S26.

$$\text{H-bond}^* = [\alpha + log(\rho_{\text{H-bond}})] \times \rho_f / \text{Vf}$$

where α is a universal parameter to avoid nonphysical results (e.g., negative values for H-bond*), $\rho_{\text{H-bond}}$ represents the averaged number density of hydrogen bonds within the framework. $\rho_f$ represents the gravimetric density of the framework, and Vf is the void fraction of the material.

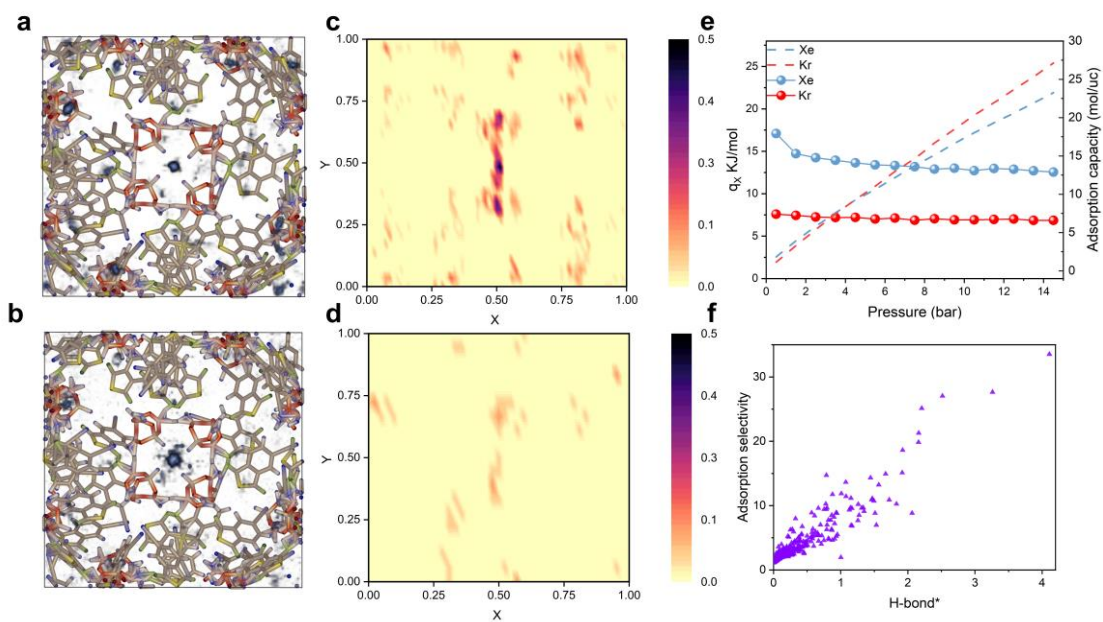

**Figure 5.** Mechanistic insights into adsorption, site analysis, and descriptor optimization for selectivity prediction. sxt+C6_NH3+L39_CN_F material a. Xe adsorption density distribution map (The isosurface is set at $10^{-5}$ mol/L); b. Kr adsorption density distribution map; c. Hydrogen bonding distribution in the HOF material; d. Hydrogen bonding distribution density in the corresponding COF framework; e. Heat of adsorption (sphere) and adsorption density (dashed line) with pressure variation (Xe:Kr=20:80, T=296 K) in the HOF; f. Correlation between adsorption selectivity and H-bond*.

## Mechanism validation with experimental data

In addition to the theoretical database, we have also comprehensively evaluated HOF materials reported in recent experimental literature. As anticipated, through feature analysis with available experimental data, we validated the universality of our proposed descriptor, which exhibits a strong linear correlation with adsorption selectivity as shown in Figure 6a. The separation mechanism was also verified with a typical experimental HOF material, as highlighted in Figure 6a. Figures 6c and 6d present the adsorption density maps for Xe and Kr, respectively. The hydrogen bonding distribution map in Figure 6c revealed that the adsorbed gas molecules are predominantly located within the central-cage region, surrounded by hydrogen-bond networks, which is consistent with the mechanism proposed in this work. Guided by the universal descriptor with enhanced mechanism understanding, we suggest that experimentalists increase framework density while maintain porosity and incorporate functional groups to enhance effective hydrogen bonding density, to realize rational design of high-performance HOF materials for gas separation.

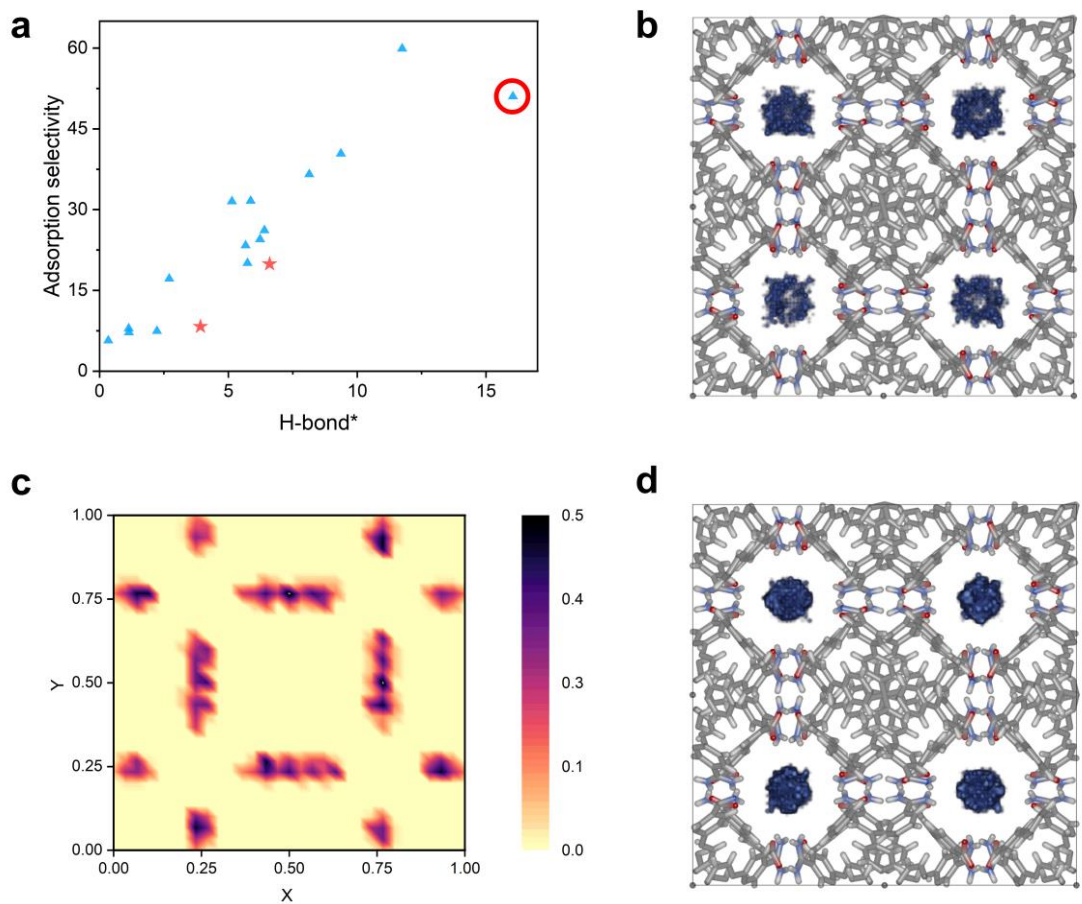

**Figure 6.** Experimental validation on separation mechanism. (a) Relationship between the proposed descriptor and adsorption selectivity from experimentally synthesized HOFs, where stars represent experimentally measured selectivity data (HIAM-103[39] and HOF-FJU-46a[1]), and the triangles are computed selectivity with experimentally synthesized HOFs (Xe/Kr selectivity has not been measured). (b) Adsorption density map of Kr gas in the structure of Bbiphen (2D)[12]. (c) Hydrogen bonding arrangement in the framework. (d) Adsorption density map of Xe gas in the structure of Bbiphen (2D)[12], where Bbiphen (2D) corresponds to the circled structure in Figure 6(a).

## Discussion

In this work, we have successfully demonstrated a hypothetical framework to construct the first large-scale HOF database with a material genomics method and integrate active learning into the inverse design workflow to achieve efficient iteration and optimization towards target materials. More importantly, we have leveraged inverse design tools to elucidate the underlying design principles of high-performance HOF materials, thereby revealing key mechanisms governing gas separation performance. Utilizing this tool, we have successfully designed a novel HOF structure with superior Xe/Kr separation selectivity up to $10^3$, while simultaneously maintaining a high Xe adsorption capacity. After comprehensive evaluation of the separation mechanism of Xe/Kr in both theoretical and experimental databases, a novel universal descriptor was established to enable rapid performance prediction bypassing cumbersome molecular simulations, with the screening speed enhanced by at least two orders of magnitude. With continuous advancement in modular manipulation strategies of HOFs, the material design concept we proposed in this work holds great promise for application in automated experimental laboratories. Furthermore, the iterative inverse engineering framework can be generalized to various scenarios in promoting systematic mechanism investigation and achieving rational design of novel materials.

## ACKNOWLEDGMENT

This work was supported by National Natural Science Foundation of China (No. 22108256), Natural Science Foundation of Henan Province (No. 252300421176) and National Natural Science Foundation of China (No. 22478361). The computations were performed at National Supercomputing Center in Zhengzhou, China.

## DATA AVAILABILITY STATEMENT

The raw data and source code for the article are available from the corresponding author upon reasonable request.

**Supplementary Information**

**A Mechanism-Guided Inverse Engineering Framework to Unlock Design Principles of H-Bonded Organic Frameworks for Gas Separation**

Yong Qiu[1], Lei Wang[1], Letian Chen[2], Yun Tian[1]*, Zhen Zhou[1,2]*, Jianzhong Wu[3]

[1]Interdisciplinary Research Center for Sustainable Energy Science and Engineering (IRC4SE²), School of Chemical Engineering, Zhengzhou University, Zhengzhou 450001, Henan, China

[2]School of Materials Science and Engineering, Institute of New Energy Material Chemistry, Renewable Energy Conversion and Storage Center (ReCast), Key Laboratory of Advanced Energy Materials Chemistry (Ministry of Education), Nankai University, Tianjin 300350, China

[3]Department of Chemical and Environmental Engineering, University of California, Riverside, CA, 92521, USA

* Corresponding author. E-mail: ytian009@zzu.edu.cn; zhouzhen@nankai.edu.cn

## METHODS

### Molecular models

In this study, we focus on the calculation of adsorption capacity for a gas mixture of Xe/Kr at 296 K and 1 bar, which is the typical industrial separation condition.[1, 2, 3] The Henry's constant is evaluated at 298 K, and we employ the high-throughput computational method to construct the largest HOF databases. To improve computational efficiency, all HOF structures are treated as rigid bodies.[4] Structural features are calculated using Zeo++ software[5], while hydrogen bonding statistics are computed using VMD (Visual Molecular Dynamics)[6] and the adsorption density map was generated using iRASPA[7].

The interatomic forces are described by the Lennard-Jones (LJ) model potential, given the non-polar nature of Xe and Kr. HOFs are parameterized with the universal force field (UFF)[8] and the Optimized Potentials for Liquid Simulations force fields (OPLS)[9], for all nonbonded interactions. The detailed force field parameters for gas molecules can be found in Table S2. Detailed methodology explanation on CDFT can be found in our previous work.[10] The 12-6 LJ potential is truncated and shifted to zero at a distance of 12.9 Å.

**Adsorption separation and membrane separation**

In the low-pressure limit, the ideal adsorption selectivity can be considered as the ratio of Henry's constants[11].

$$\alpha^{IM} = \frac{K_{h,2}}{K_{h,1}} \tag{1}$$

where $K_{h,i}$ represents the Henry's constant of component i. For a gas molecule with a fixed conformation, the Henry's constant can be determined by integrating the external potential with respect to its interaction with the nanoporous material.[12]

$$K_h = \frac{1}{8\pi^2 k_B T V} \int d\varpi \int d r \, exp[-\beta \phi^{ext}(r,\varpi)] \tag{2}$$

where $\beta = 1/(k_B T)$, $k_B$ stands for the Boltzmann constant, T is the absolute temperature, V represents the system volume, and $\phi^{ext}$ is the external potential.

The actual adsorption selectivity ($\alpha$) is the ratio of the amount adsorbed to the ratio of the adsorbate molecules in the bulk. In this work, the bulk ratio $y_{Xe}:y_{Kr} = 2:8$.

$$\alpha = \frac{\Gamma_2/y_2}{\Gamma_1/y_1} \qquad (3)$$

$\Gamma_i$ is the adsorption capacity, and $y_i$ is the bulk molar fraction.

**DFT calculations**

The density functional theory (DFT) computations with van der Waals (vdW) corrections were carried out in the Vienna ab initio Simulation Package (VASP).[13] The Perdew-Burke-Ernzerhof (PBE) functional within the generalized gradient approximation (GGA) was used to describe the exchange-correlation interaction.[14] Projector augmented wave (PAW) methods were used for pseudopotentials.[15] An energy cutoff of 400 eV was adopted for the plane-wave basis set. A vacuum layer of 15 Å was used to prevent the interaction between periodic images. The energy convergence criterion for geometric optimization is set to be $10^{-6}$ eV. The Γ-centered Monkhorst-Pack scheme with a K-point resolution of 0.03 π/Å was used in all DFT computations. The DFT-D3 method was adopted to account for the van der Waals interaction.[16] The VASPKIT was utilized for post-processing of computational data.[17]

The adsorption energy ($E_{ads}$) was calculated according to the equation below,

$$E_{ads} = E_{12} - E_1 - E_2 \qquad (4)$$

where $E_{12}$ is the total optimized energy of the adsorbed structure, $E_1$, $E_2$ are the optimized energies of the adsorbate and adsorbent.

Charge density difference is defined as the total charge density of the system subtracted by the charge density of the adsorbed molecules on the substrate. The rigorous definition is as follows,

$$\Delta\rho = \rho_{A/S} - \rho_A - \rho_S \qquad (5)$$

where $\rho_{A/S}$ represents the total charge density of the adsorbed molecule on the substrate, with $\rho_A$ and $\rho_S$ denoting the charge densities of the adsorbed molecule and the substrate, respectively.

## I. Supplementary Tables

**Table S1.** Abbreviations and their full names

| Abbreviation | Full name |
|---|---|
| GLD | Largest included sphere diameter |
| LCD | Largest free sphere diameter |
| PLD | Largest included sphere along free sphere path diameter |
| density | Framework density |
| VSA | gravimetric surface |
| GSA | volumetric surface |
| Vp | Void proce-occupiable volume fraction |
| Vf | Void fraction |
| Herry_Kr | Henry's constant of Kr |
| Herry_Xe | Henry's constant of Xe |
| N_H | Num hydrogen |
| N_C | Num carbon |
| N_N | Num nitrogen |
| N_F | Num fluorine |
| N_Cl | Num chlorine |
| N_Br | Num bromine |
| N_I | Num iodine |
| TDU | Total degree of unsaturation |
| ETR | Electronegative atoms to total atoms ratio |
| α | Adsorption selectivity |
| LASSO | Least Absolute Shrinkage and Selection Operator |
| SVM | Support Vector Machine |
| PLS | Partial Least Squares |
| RF | Random Forest |
| GBR | Gradient Boosting Regression |
| ABR | Adaptive Boosting Regression |
| ANN | Artificial Neural Network |
| KNN | K-Nearest Neighbors |
| SISSO | Sure Independence Screening and Sparsifying Operator |
| SHAP | SHapley Additive exPlanations |
| $N_{H-bond}$ | Number of hydrogen bonds |

**Table S2.** The Lennard-Jones (LJ) parameters for Xe and Kr.

| Molecule | σ (Å) | ε/kB (K) |
|----------|-------|----------|
| Xe       | 4.10  | 211.0    |
| Kr       | 3.68  | 110.0    |

## II. Supplementary Figures

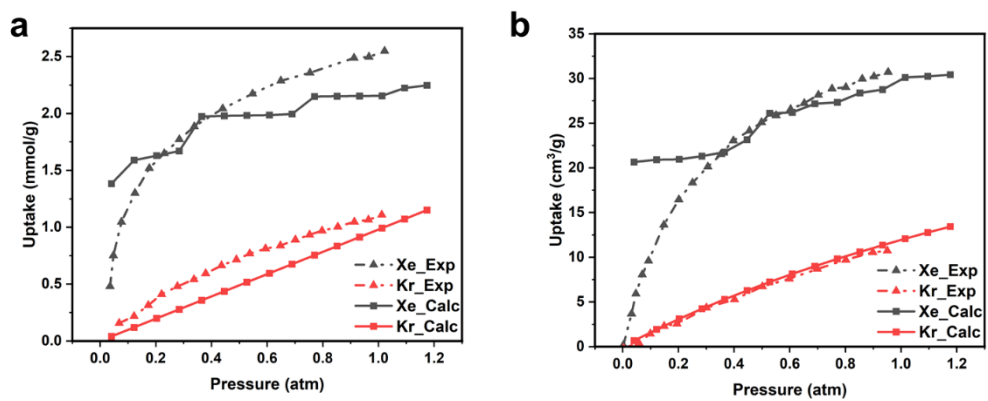

**Figure S1.** Comparison of CDFT calculations with experimental data (a[18], b[1]). Triangular markers represent experimental data, while square markers correspond to theoretical predictions.

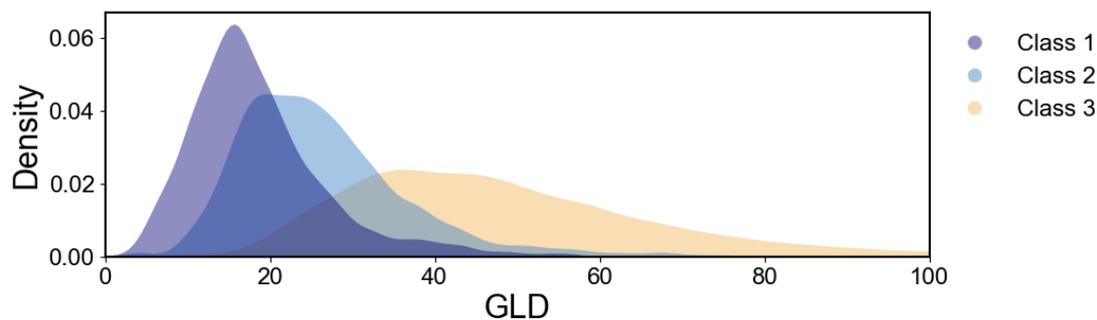

**Figure S2.** Distribution of GLD in the database.

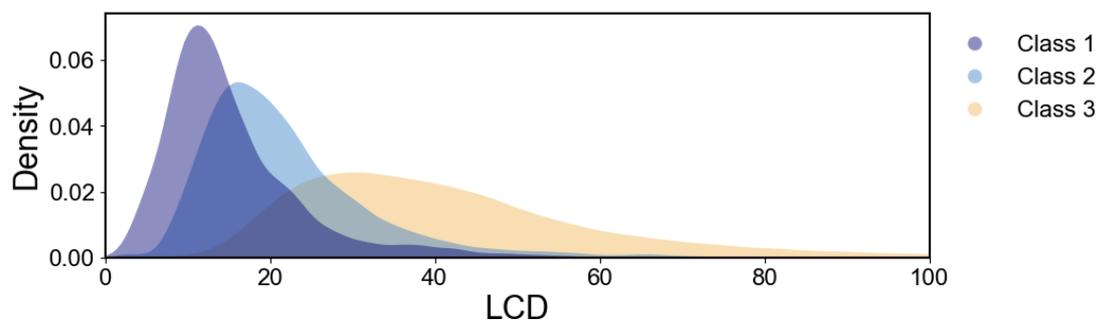

**Figure S3.** Distribution of LCD in the database.

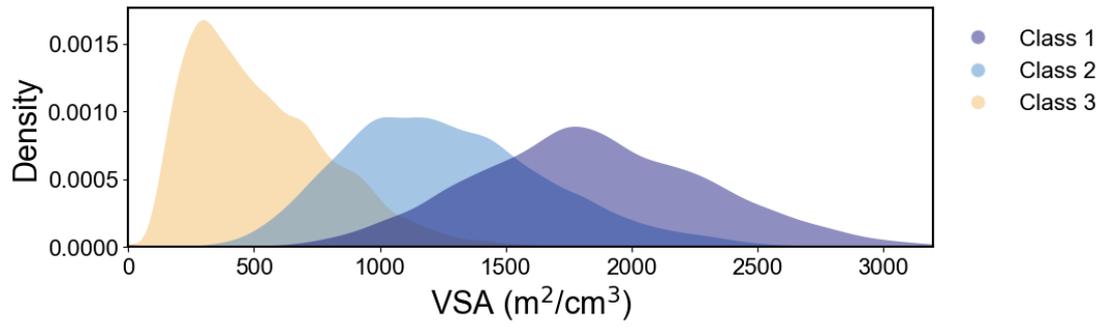

**Figure S4.** Distribution of VSA in the database.

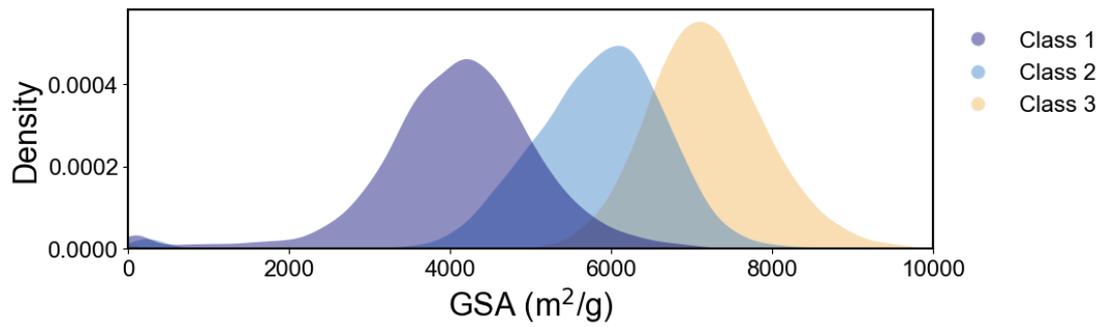

**Figure S5.** Distribution of GSA in the database.

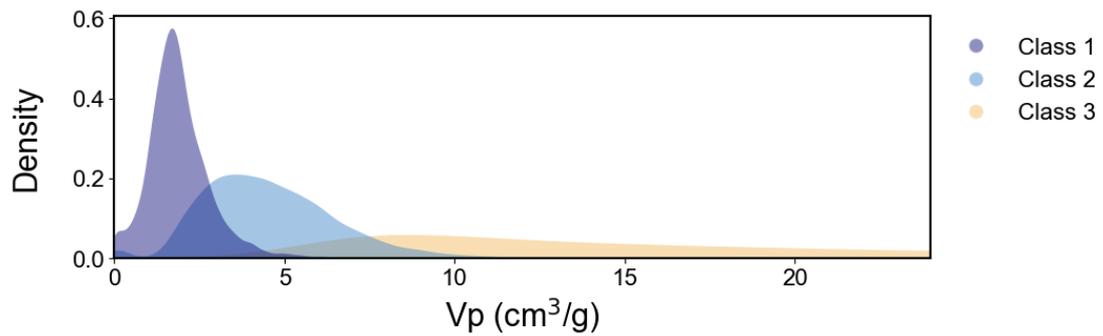

**Figure S6.** Distribution of Vp in the database.

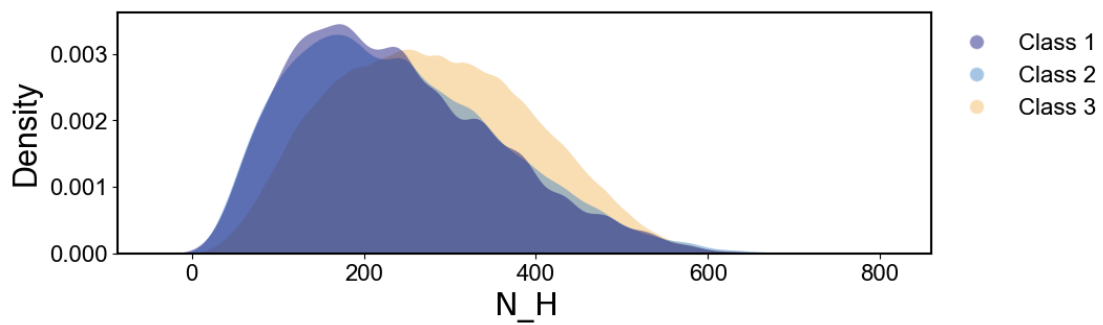

**Figure S7.** Distribution of the number of H in the database.

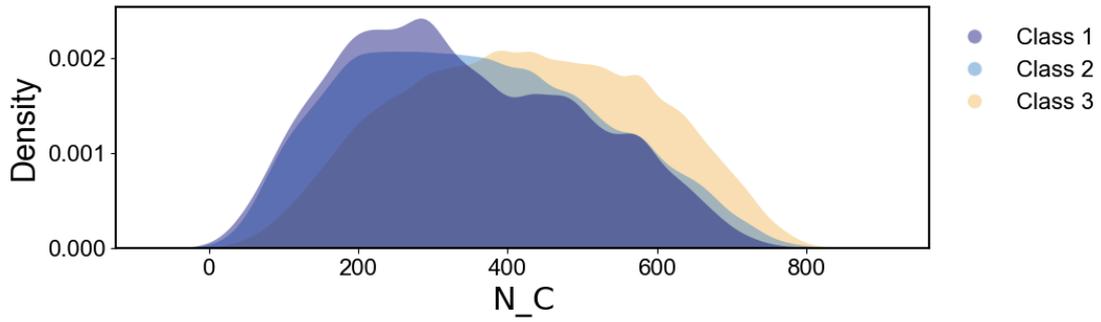

**Figure S8.** Distribution of the number of C in the database.

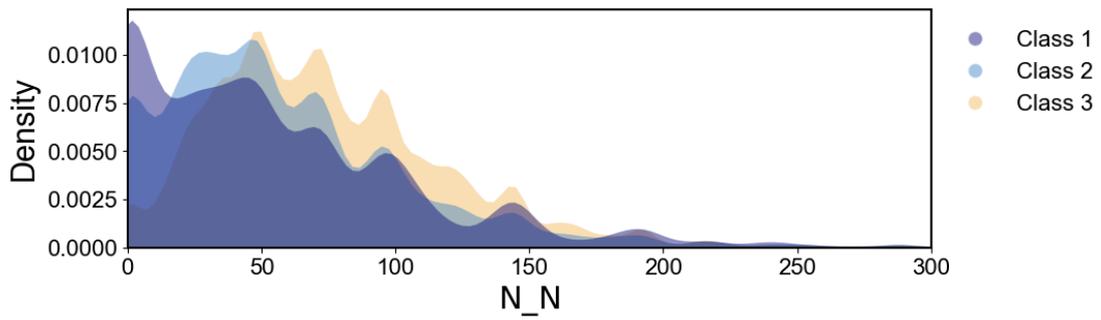

**Figure S9.** Distribution of the number of N in the database.

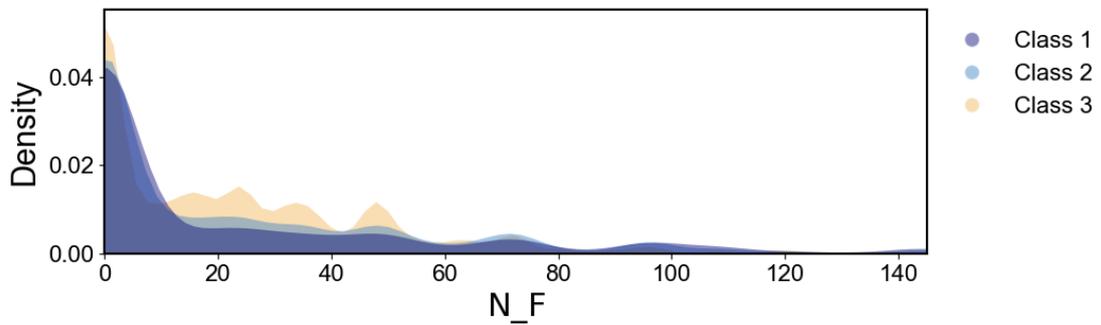

**Figure S10.** Distribution of the number of F in the database.

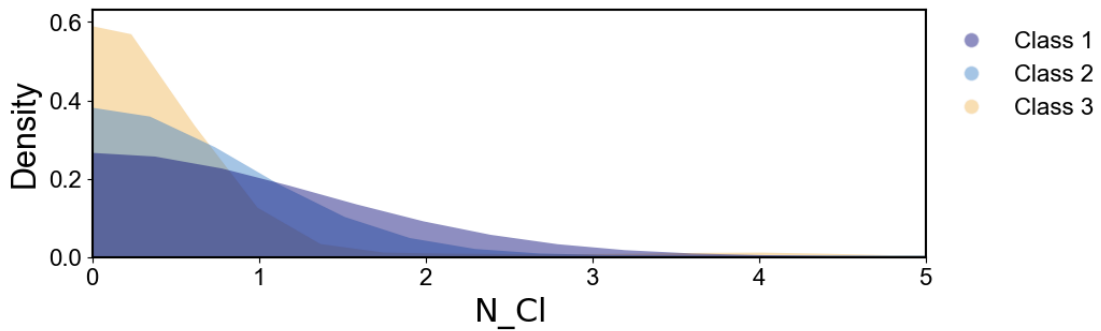

**Figure S11.** Distribution of the number of Cl in the database.

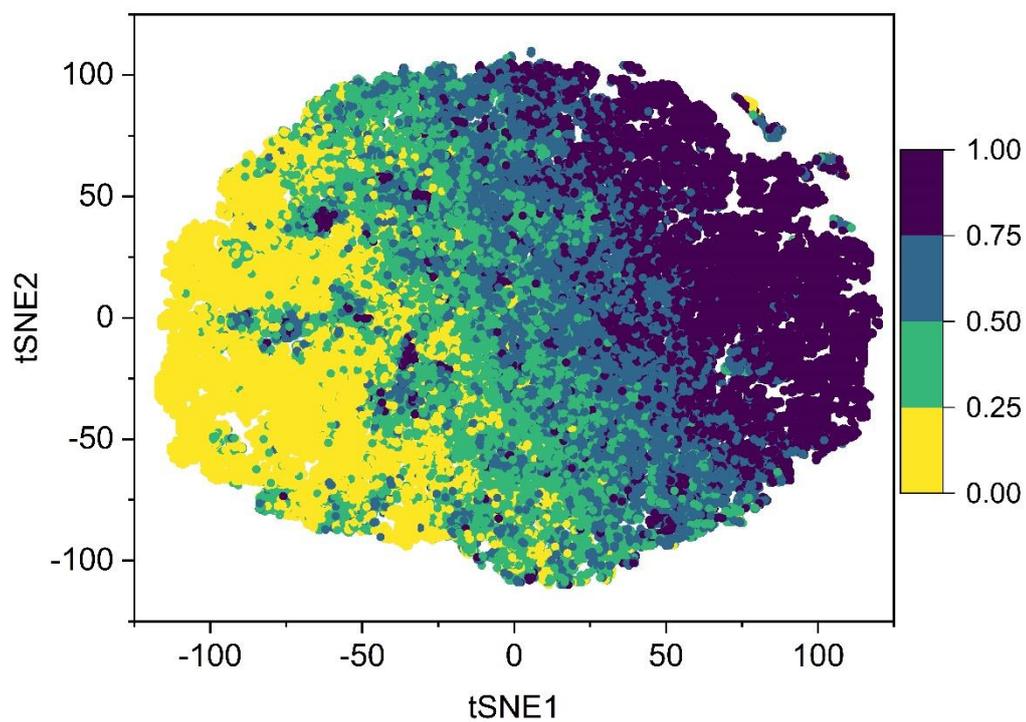

**Figure S12.** t-SNE distribution plot, with color mapping representing the adsorption selectivity. The darker the color, the higher the selectivity.

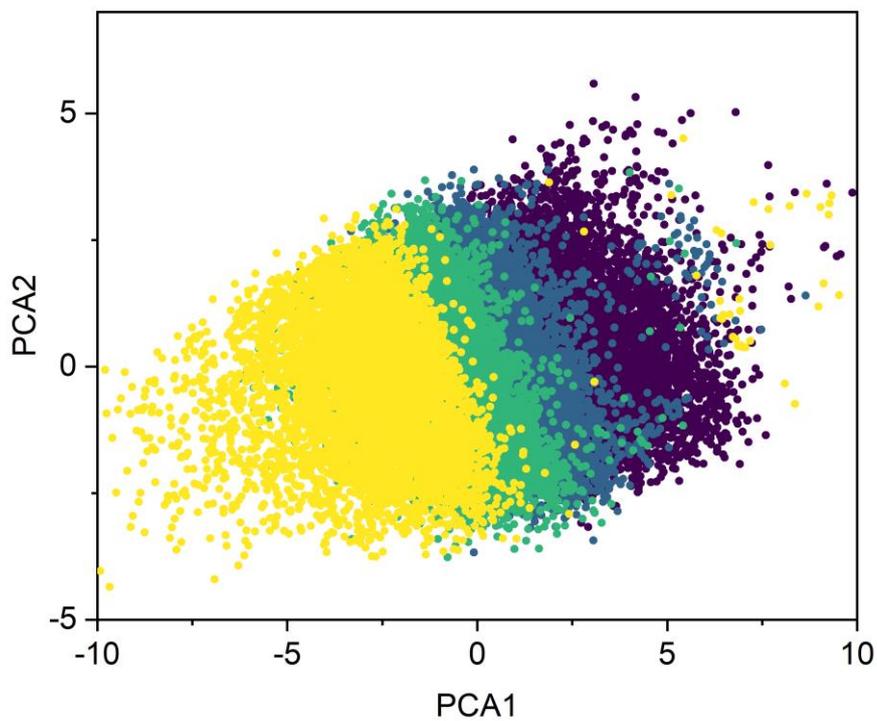

**Figure S13.** PCA distribution plot, with color mapping representing the adsorption selectivity. The darker the color, the higher the selectivity.

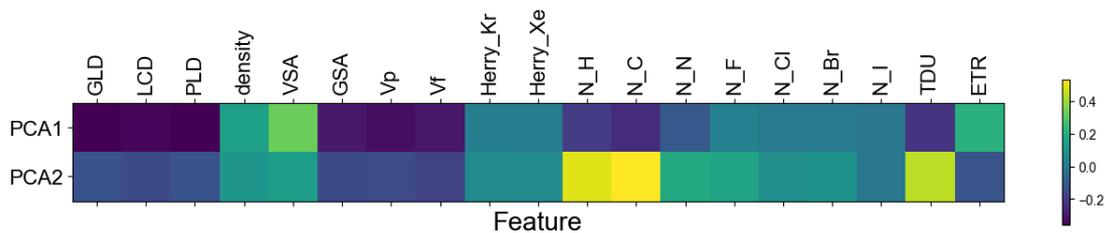

**Figure S14.** Weight distribution plot of each principal component in PCA, with color mapping representing the magnitude of the weights.

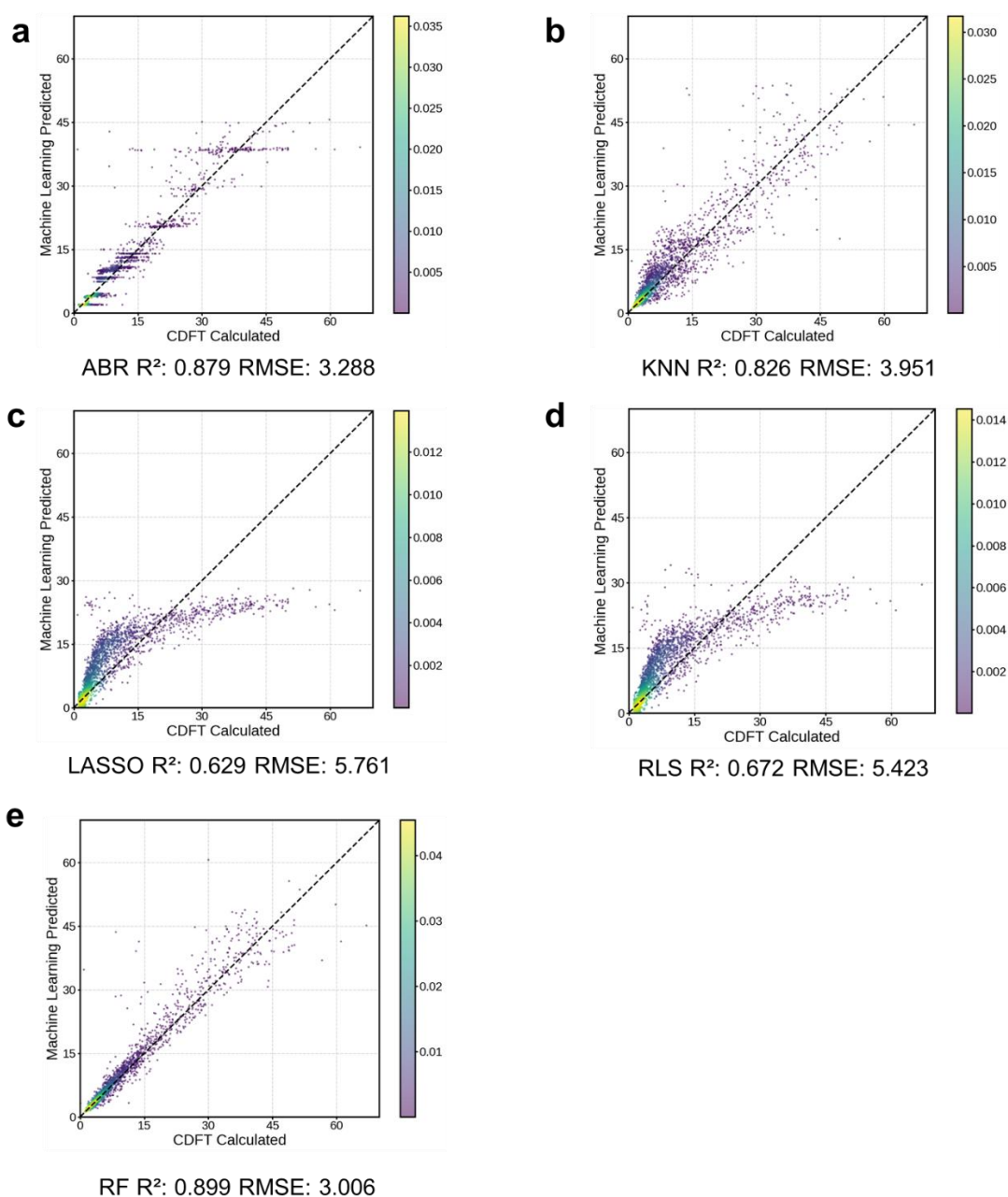

**Figure S15.** Performance of various machine learning methods. a. ABR b. KNN c. LASSO d. RLS e. RF. The SVM and ANN algorithms are not shown due to large deviations.

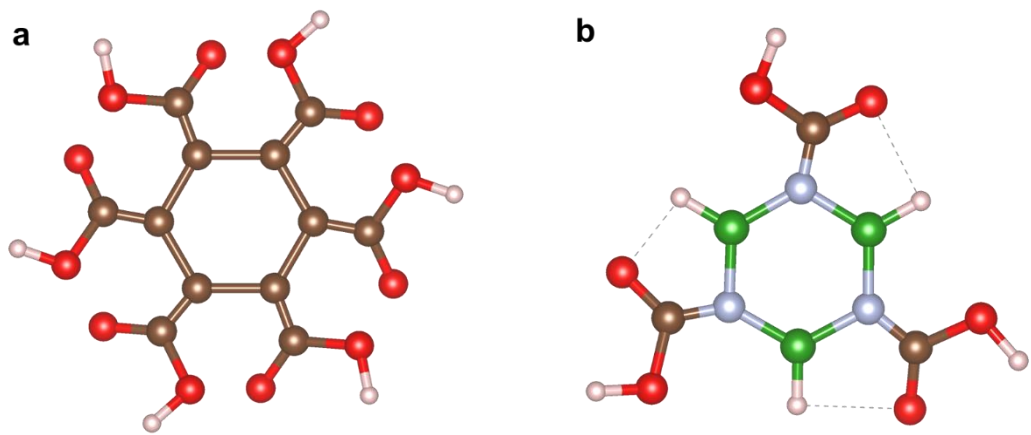

**Figure S16.** Comparison of node building blocks: a. C36_COOH, b. C30_COOH, the former matched with more ligands to form additional hydrogen bonds.

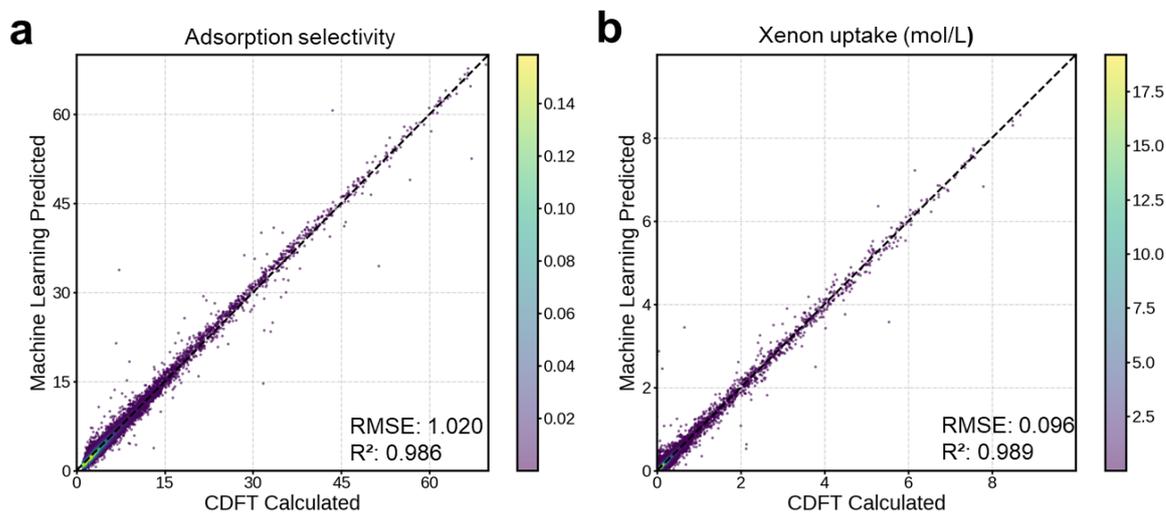

**Figure S17.** Accuracy of the MOF-NET model for adsorption selectivity (a) and Xe adsorption capacity (b).

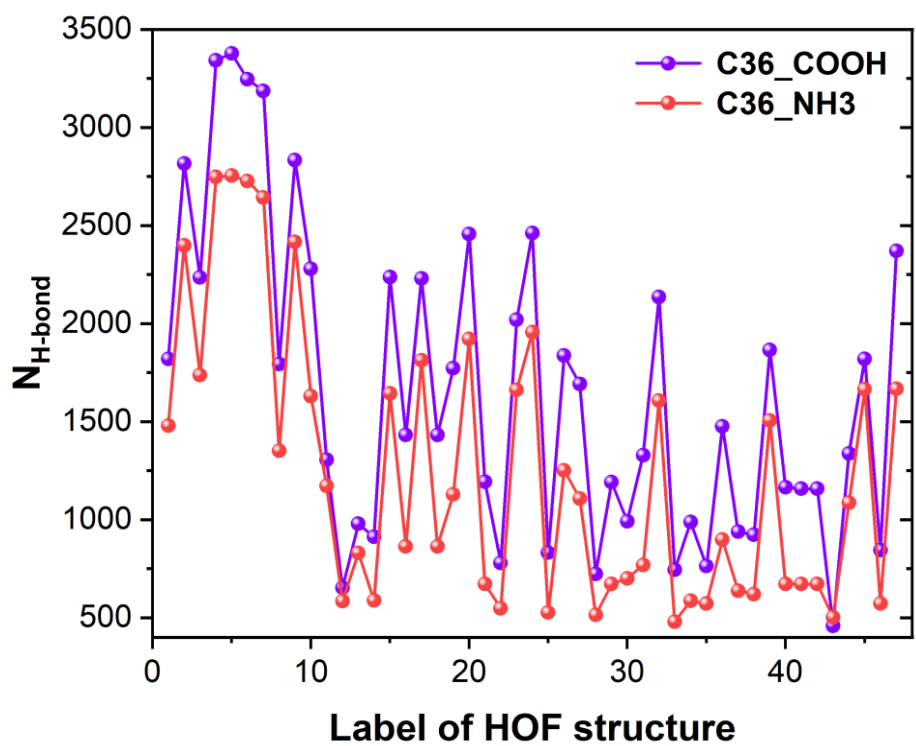

**Figure 18.** The number of hydrogen bonds corresponding to CIF structures with only the node being displaced.

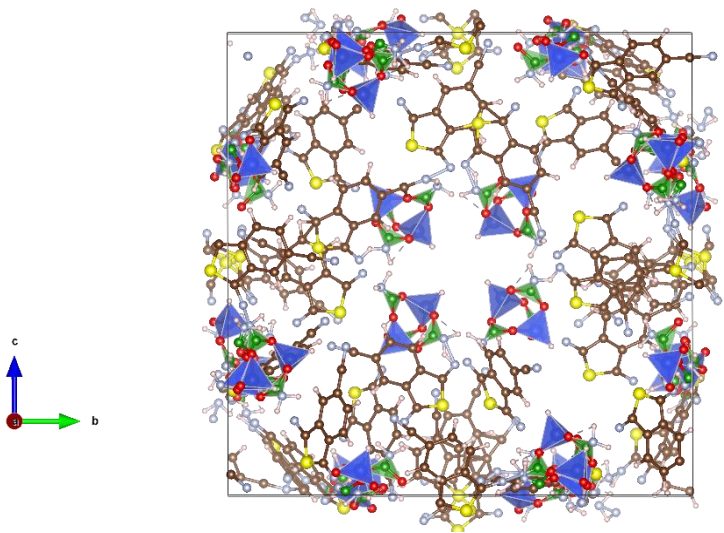

**Figure S19.** sxt+C6_NH3+L39_CN_F schematic diagram.

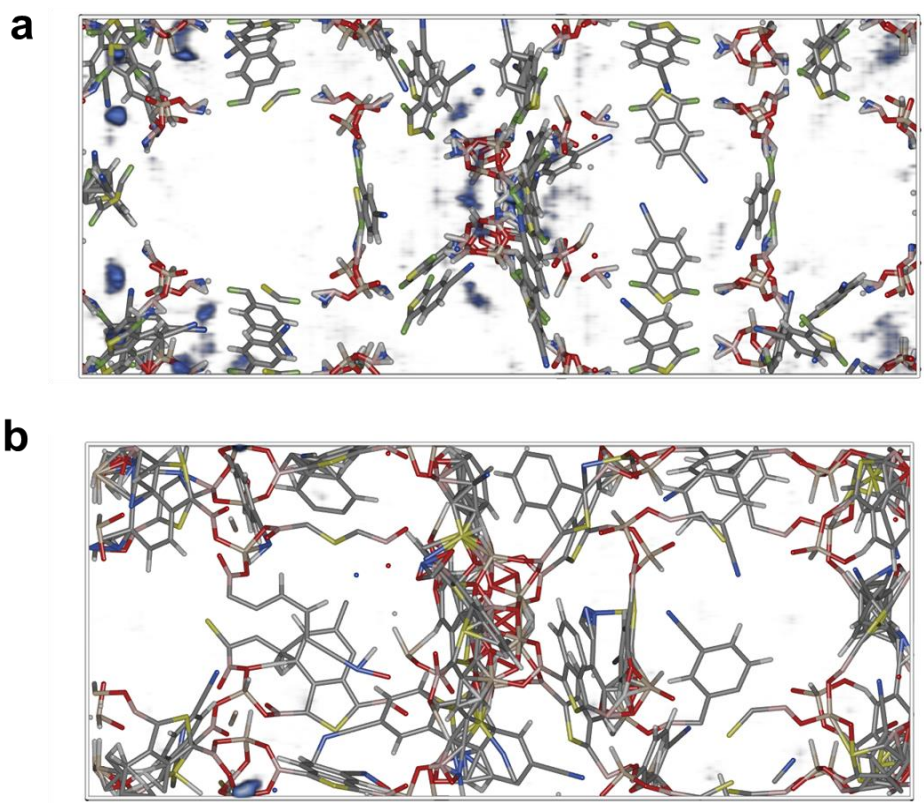

**Figure S20.** adsorption density distribution map of sxt+C6_NH3+L39_CN_F(a)( XY direction) and sxt+C6+L39_CN (b).(The isosurface is set at $10^{-5}$ mol/L)

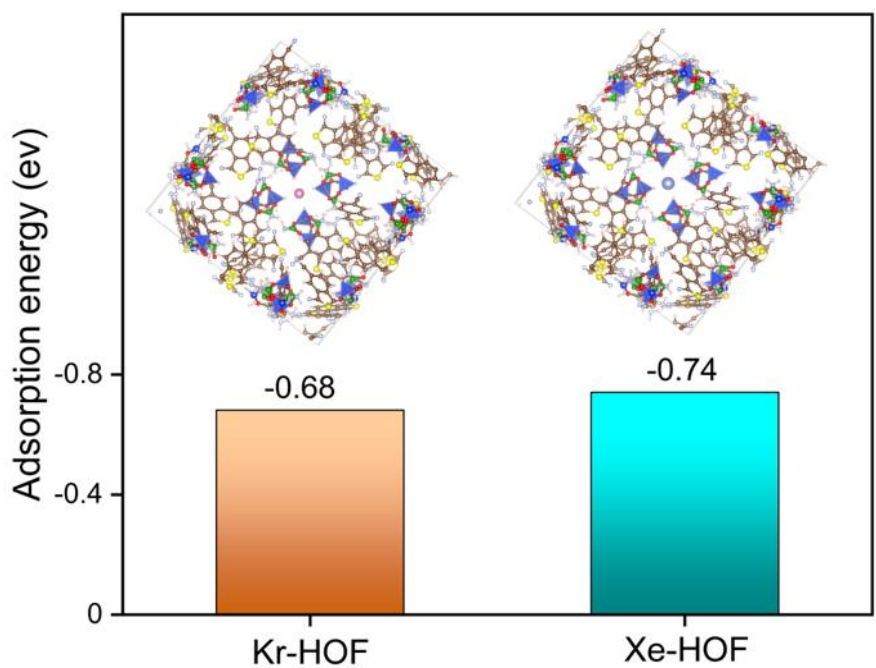

**Figure S21.** The adsorption energies calculated using DFT for Xe and Kr molecules.

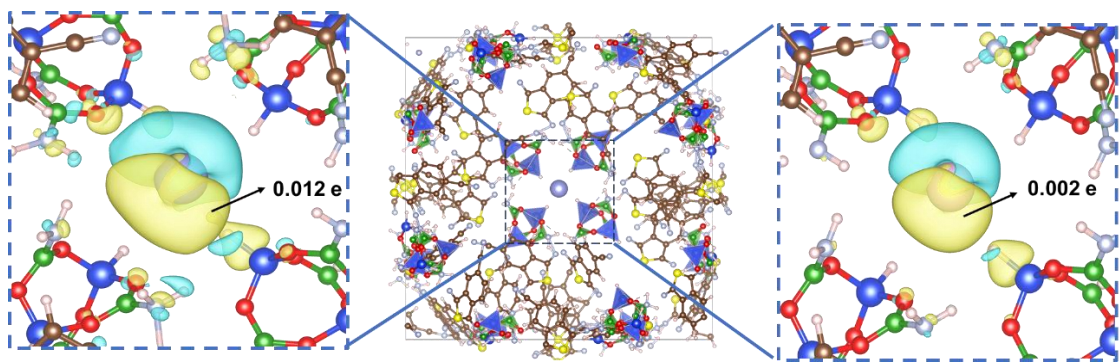

**Figure S22.** The differential charge distribution of guest molecule adsorption in HOF, with Xe on the left and Kr on the right. (The isosurface for sxt+C6_NH3+L39_CN_F is set at 0.0001 e/Bohr³)

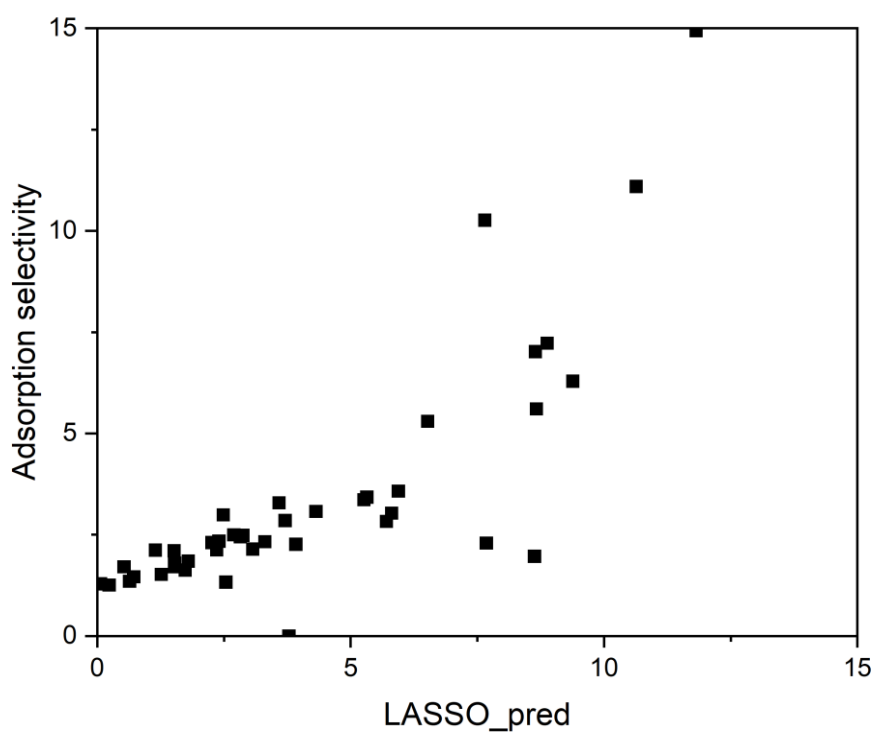

**Figure S23.** Equation fitting of Lasso prediction. (LASSO_pred= 0.01061*GLD+ 0.0040853*VSA+ 0.003873*Herry_Kr+ 0.0019689*$N_{\text{H-bond}}$).

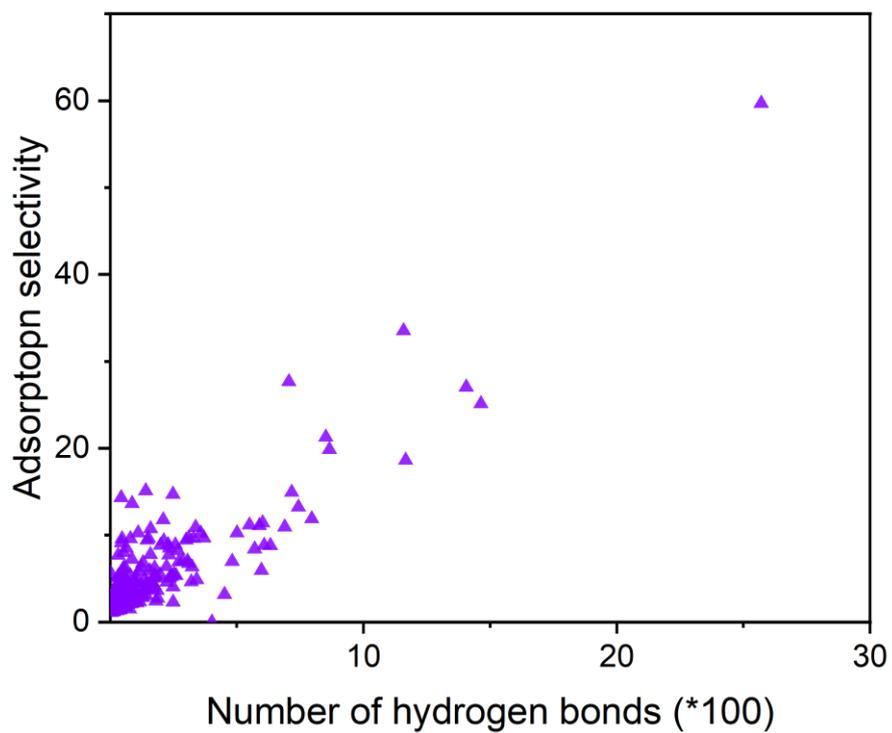

**Figure S24.** The relationship between number of hydrogen bonds and adsorption selectivity.

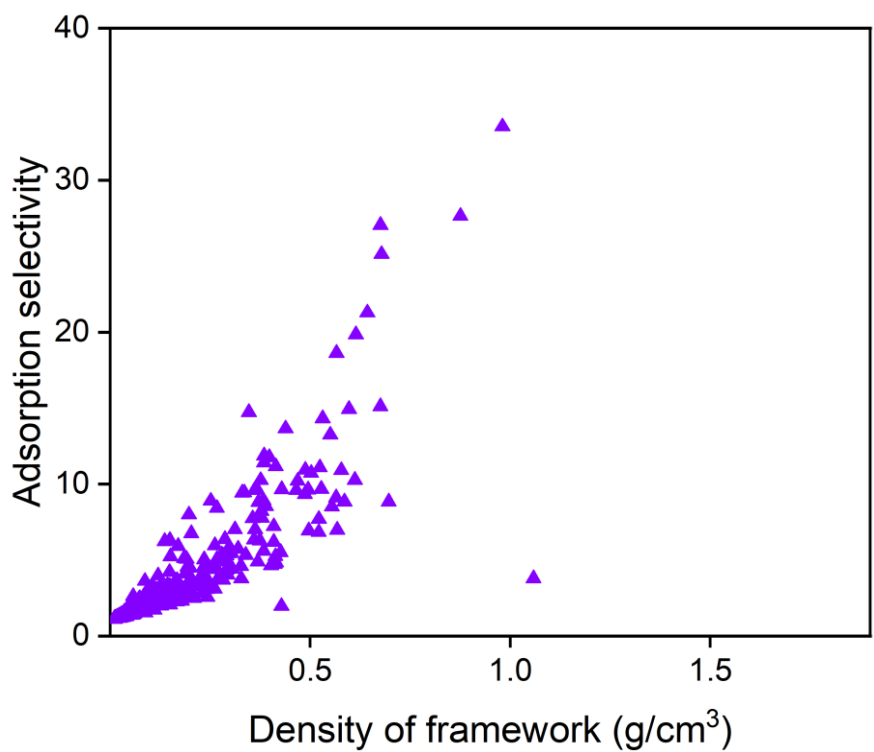

**Figure S25.** The relationship between density of framework and adsorption selectivity.

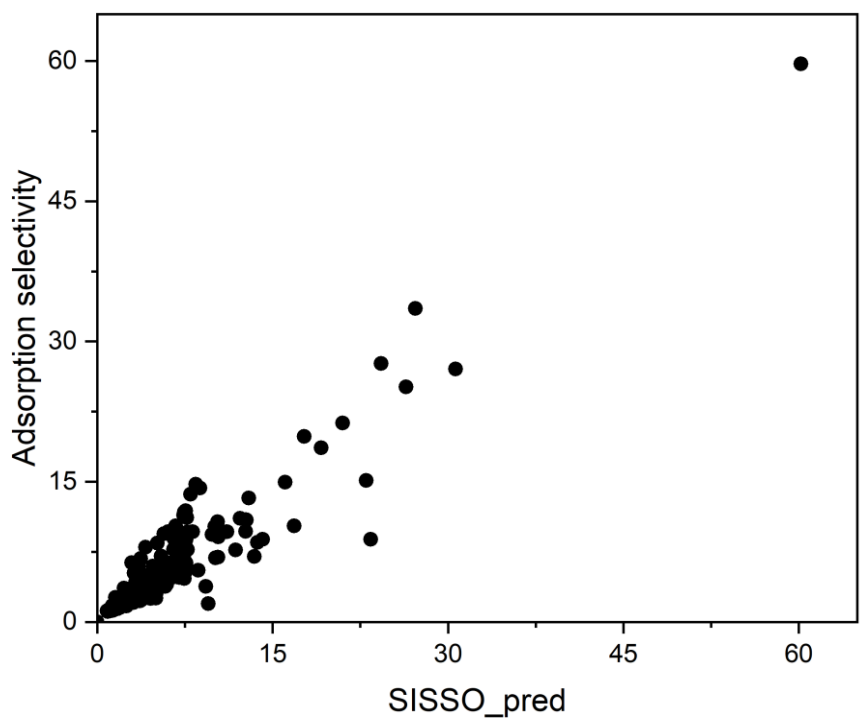

**Figure S26.** Equation fitting of SISSO prediction. (SISSO_pred=sum(28.99510943*(exp(-(Vp)^3))-31.9562961* (abs(density - void_fraction)+ density)+ 0.0136838947* ((density)^3* $N_{H-bond}$)+ 32.61811921)